\documentclass[aps,prb,reprint,pacs,groupaddress]{revtex4-1}

\usepackage{graphicx}

\usepackage{dcolumn}
\usepackage{bm}
\usepackage{amsmath,amssymb}

\renewcommand{\d}{{\rm d}}
\newcommand{\e}{{\rm e}}
\newcommand{\imai}{{\rm i}}

\usepackage{color}

\begin{document}

\title{Unraveling of free carrier absorption for terahertz radiation in
heterostructures}

\author{Andreas Wacker}
\email[]{Andreas.Wacker@fysik.lu.se}
\affiliation{Mathematical Physics, Lund University, Box 118, 22100 Lund, Sweden}
\author{Gerald Bastard}
\author{Francesca Carosella}
\author{Robson Ferreira}
\affiliation{Laboratoire Pierre Aigrain, Ecole Normale Sup{\e}rieure, CNRS
  (UMR 8551),
\\
Universit{\e} P. et M. Curie, Universit{\e} D. Diderot, 
24 rue Lhomond F-75005 Paris, France}
\author{Emmanuel Dupont}
\affiliation{Institute for Microstructural Sciences, National Research Council, Ottawa, Ontario, Canada K1A0R6}

\date{2. November 2011, accepted by Physical Review B}

\begin{abstract} 
The relation between free carrier absorption and intersubband
transitions in semiconductor heterostructures is resolved by comparing a
sequence of structures. Our numerical and analytical results show how
free
carrier absorption evolves from the intersubband transitions in
the limit of an infinite number of wells with vanishing barrier width. It is
  explicitly shown that the integral of the absorption over frequency
matches the value obtained
by the f-sum rule. This shows that a
proper treatment of intersubband transitions is fully sufficient to simulate
the entire electronic absorption in heterostructure THz devices.
\end{abstract}


\maketitle

\section{Introduction}\label{SecIntro}
The absorption of electromagnetic radiation due to the  interaction
with electrons in bulk crystals is essentially determined by two
distinct effects: (i) The free carrier absorption (FCA), which is directly
related to the  electrical conductivity and drops with frequency on
the scale of the inverse scattering time. (ii) Interband transitions, which
are typically described via the dipole moments induced by the coupling
between states in different bands. For most crystals these transition
energies are of the order of eV and thus this dominates the 
response around the optical spectrum. In addition to these electronic 
features, optical phonons provide absorption in the far infrared region, which
is not addressed here. 

Semiconductor heterostructures provide an additional effective
  potential for the electron in the conduction band causing a further
 quantization of the electronic states in the growth direction (denoted by
 $z$). Taking into account the degrees of freedom for motion in the $x,y$
 plane, this establishes subbands within the conduction band. Commonly, the
absorption between these subbands is treated analogously to the interband
transitions in bulk crystals. The standard treatment relies
on the envelope functions $\varphi_\nu(z)$ for the subbands $\nu$ with energies
$E_\nu$ and areal electron densities $n_\nu$ using expressions for the
absorption coefficient $\alpha_{\mu\to\nu}(\omega)$ as
\cite{HelmReview1999,AndoJPSJ1978} 
\begin{equation}\begin{split}
\alpha_{\mu\to\nu}(\omega)=&\frac{e^2|z_{\mu\nu}|^2(E_\nu-E_\mu)(n_\mu-n_\nu)}{
2\hbar L_zc\sqrt{\epsilon}\epsilon_0}\\
&\times\frac{\Gamma}{(E_\nu-E_\mu-\hbar\omega)^2+\Gamma^2/4}
\label{EqAlphaIntersubband}
\end{split}\end{equation} 
where $E_\mu<E_\nu$ and
the counter-rotating terms are neglected. Here
$e$ is the elementary charge, $\sqrt{\epsilon}$ is the refractive index,
and $\epsilon_0$ is the vacuum
permeability (SI units are used).
The matrix element
\begin{equation}
z_{\mu\nu}=\int \d z \varphi_\mu(z)z \varphi_\nu(z)\label{Eqzmatrix}
\end{equation} 
describes the coupling strength. Throughout this work  we assume the
polarization of the electric field to point in $z$-direction 
and that the wave propagates in a waveguide of effective 
thickness $L_z$ which is filled by the (layered) semiconductor material. This
scheme is also routinely applied for the calculation of the gain
spectrum of quantum cascade lasers (QCLs).\cite{FaistScience1994} In this
context the broadening $\Gamma$ can be either added in a
phenomenological way \cite{AjiliJAP2006} or by detailed calculations,
see, e.g., Ref.~\onlinecite{UnumaJAP2003}. It can also be seen as a 
limiting case of a full quantum kinetic calculation.\cite{WackerSPIE2009}

While the conventional treatment of intersubband transitions is well accepted
for transitions in the infrared, this approach is less obvious for THz
systems, which have become of high
interest.\cite{WilliamsNatPh2007,LeeScience2007} Here, FCA-related features
might turn up as a strong competing mechanism to the intersubband gain
transition in analogy to the bulk case where both FCA and interband
transitions occur as separate processes. In order to demonstrate the potential
relevance, we consider the standard expression
for FCA in bulk systems\cite{YuBook1999} 
\begin{equation}
\alpha_\mathrm{FCA}(\omega)=\frac{n_c e^2\tau}{m_cc\sqrt{\epsilon}\epsilon_0}
\frac{1}{\omega^2\tau^2+1}
\label{EqDrude}
\end{equation} 
where $m_c$  is the effective mass,
$n_c$ the volume density of electrons in the conduction
band, and  $\tau$ is the scattering time. As an
example for GaAs with a doping of $1\times 10^{16}/\textrm{cm}^3$ and
$\tau=0.2$ ps (corresponding to a mobility of 6000 cm$^2$/Vs at 300K
\cite{MeyerPRB1987}) one obtains $\alpha=120/\textrm{cm}$ for a frequency
$\omega/2\pi=2$ THz. This is larger than typical gain coefficients in THz
quantum cascade lasers.\cite{JukamAPL2009,MartlOE2011,BurghoffAPL2011} Thus,
bulk FCA would provide a strong obstacle in achieving lasing in such
structures and its proper treatment in heterostructures is of crucial
importance for the description of QCLs or other THz heterostructure
devices. (For a typical infrared laser, in contrast, it was shown that FCA in
the cascade structure  does not play a role.\cite{GiehlerJAP2004}) In
Ref.~\onlinecite{AjiliJAP2006} FCA was only considered in the waveguide layers but
not the QCL structure itself, where the absorption was determined by
intersubband transitions. Furthermore, in Ref.~\onlinecite{VurgaftmanPRB1999} it was
shown that processes as described by Eq.~(\ref{EqAlphaIntersubband}) dominate
the absorption of light (with $z$-polarized electric field) for quantum wells.

In this context the question arises how such a treatment based on
intersubband transitions is related to the FCA in the bulk. Is FCA
related to the seemingly dominating  intersubband processes or does it stem from
further processes not identified yet? In the latter case, such
processes could strongly alter the THz performance of heterostructure
devices. In order to shed light on this important issue we present a
detailed study on the unfolding of FCA starting from different types
of heterostructure. Our main conclusion is that the absorption due to
intersubband transitions evolves into the bulk FCA for vanishing barrier widths.
This shows that a proper treatment  of
intersubband transitions provides a complete description of gain and
absorption processes in heterostructure devices.

\section{From superlattice to bulk}
\begin{figure}
\includegraphics[width=\columnwidth]{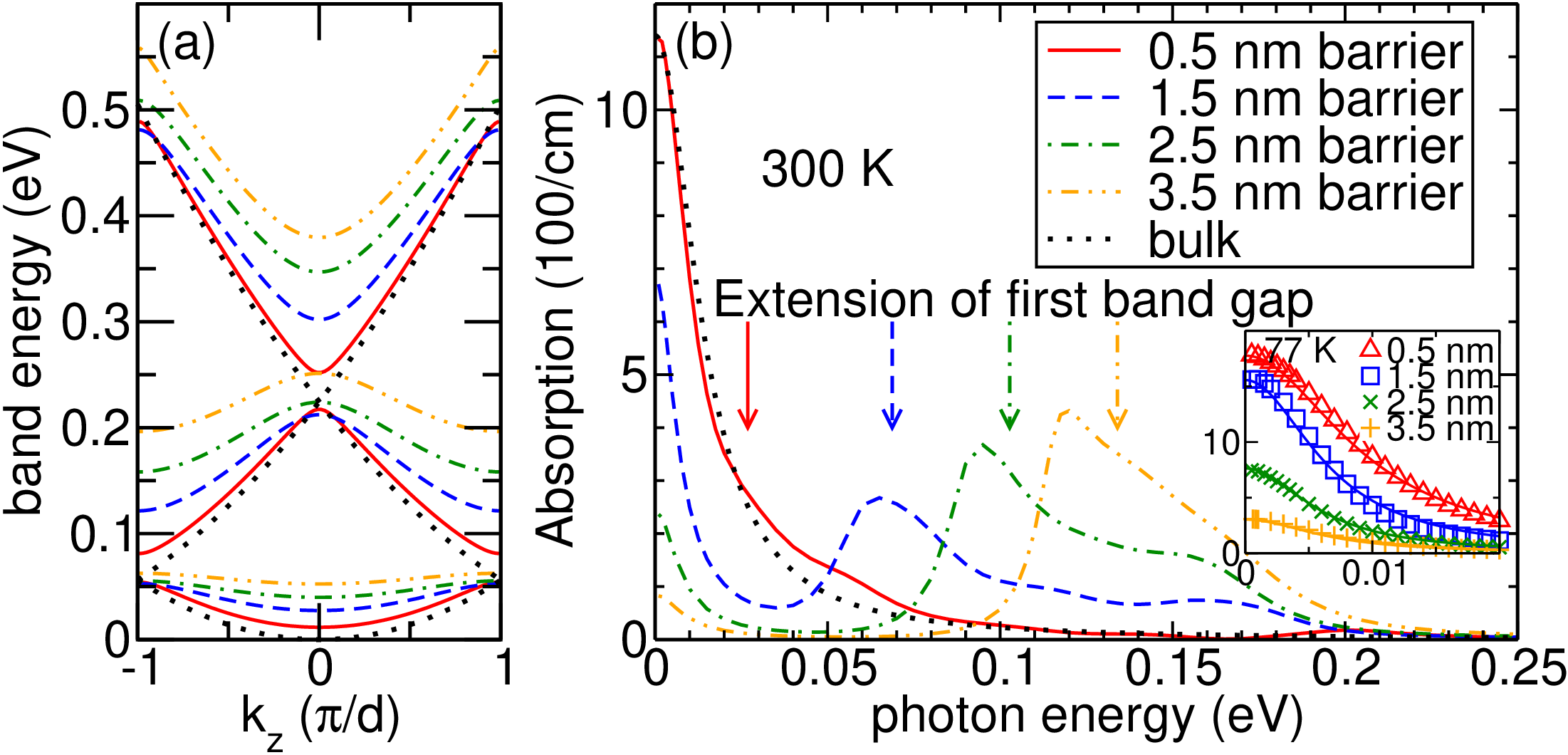}
\caption{(Color online) (a) Lowest three minibands for the
  SLs together with the dispersion of bulk GaAs (dotted line)
  neglecting nonparabolicity. (b) Absorption at 300 K for the SLs,
  calculated by the NEGF  model together with the Drude expression
  (\ref{EqDrude}) for bulk absorption (dotted line) using
  $m_c=m_\textrm{SL}=0.071m_e$ and  $\tau=46$ fs. The inset shows the
  NEGF calculations (symbols) at 77 K together with the corresponding
  result of  Eq.~(\ref{EqalphaMB}) using  $\sigma_0$ from table
  \ref{TabParameters} and $\tau_m=70/100/110/100$ fs for the sample
  with the barrier width of 0.5/1.5/2.5/3.5 nm, respectively  (lines).
\label{FigSLabs}}
\end{figure}

We consider four GaAs-Al$_{0.3}$Ga$_{0.7}$As superlattices\cite{GrahnBook1995} (SLs) with
constant period $d=10$ nm. The barrier width is set equal to  0.5
nm, 1.5 nm, 2.5 nm, and 3.5 nm, respectively, and a homogeneous doping with
$n_c=6\times 10^{16}/\textrm{cm}^3$ is used. The sample with the 2.5 nm
barrier has been investigated in Ref.~\onlinecite{HelmPRB1993}, which motivates
our choice. Fig.~\ref{FigSLabs}(a) shows the calculated minibands
assuming effective masses of $0.067m_e$  and $0.0919m_e$  for GaAs
and  Al$_{0.3}$Ga$_{0.7}$As, respectively, where $m_e$ is the 
free electron mass, as well as a conduction band offset
of  276 meV.\cite{VurgaftmanJAP2001} Further information on the
structures is given in table \ref{TabParameters}.

\begin{table}
\caption{Key parameters obtained for the different SLs. The
  effective mass $m_{SL}$ is taken for the lowest miniband at $k=0$ in the
  SL direction.}
\begin{tabular}{|l|c|c|c|c|}
\hline
barrier width (nm) & 0.5 & 1.5 & 2.5 & 3.5 \\
\hline
miniband width (meV) & 42.7 & 25.4 & 15.6 & 10.1\\
\hline
effective mass $m_{SL}/m_e$ & 0.071 & 0.090 & 0.125& 0.178\\
\hline
$\sigma_0$ at 300 K (A/Vcm) & 11.2 & 6.4 & 2.4 & 0.8\\
\hline
$\sigma_0$ at 77 K (A/Vcm) & 17 & 14.9 & 7.3 & 2.9 \\
\hline
\end{tabular}
\label{TabParameters}
\end{table}

Here the zero-field conductivity $\sigma_0$  is evaluated from the
nonequilibrium Green's function (NEGF) model following
Ref.~\onlinecite{LeePRB2006,*NelanderDiss2009}, which includes
scattering processes from phonons, impurities, interface roughness
(with an average height of one monolayer and a length correlation of
10 nm), and alloy disorder in an approximate way. This program also
calculates the absorption in linear response to the optical field
\cite{WackerSPIE2009} as given in Fig.~\ref{FigSLabs}(b). Using
$\sigma_0\approx n_ce^2\tau/m_\textrm{SL}$ the conductivity for the
0.5 nm barrier structure provides a scattering time of $\tau= 46$
fs. This value agrees roughly with the momentum scattering rate
$1/\tau_m=29/$ps (which is the sum of the elastic and
inelastic scattering rate\cite{WackerPR2002}) extracted from several
highly doped GaAs/AlAs SLs with narrow barriers at room
temperature.\cite{SchomburgPRB1998} This value is much smaller than the
bulk scattering time of 0.2 ps, as scattering is enhanced due to the
presence of rough interfaces in all SLs (which are particular strong
scatterers for small barrier widths, when the wave functions highly
penetrate through the barriers).  In addition, the assumption of a
constant scattering time is only expected to be of semi-quantitative
nature, the same holds for the approximations in matrix elements
used. (For a more detailed treatment of roughness scattering in thin
barriers, see Ref.~\onlinecite{CarosellaPRB2010}.) Using $\tau= 46$ fs,
the Drude expression (\ref{EqDrude}) fits the absorption quite well,
demonstrating, that these small barriers actually provide almost the
bulk free carrier absorption behavior. 

With increasing barrier thickness the conductivity becomes smaller due
to the reduced coupling between the quantum wells. Accordingly, there
is a decrease in the low frequency absorption
\begin{equation}
 \alpha(\omega=0)=\frac{\sigma_0}{c\sqrt{\epsilon}\epsilon_0}\, ,
\label{EqAlphaSigma}
\end{equation}
as it follows from electrodynamics.\cite{JacksonBook1998} Here our
numerical calculations are in full agreement, as we do not employ the
rotating wave approximation and include broadening in a fully
consistent way. Furthermore, for thicker barriers, the absorption
between the minibands becomes more prominent and thus the absorption
increases close to the photon energy required to overcome the gap
between the first and the second miniband, as indicated by the arrows
in Fig.~\ref{FigSLabs}(b). The shift of the peak positions with
respect to the minigaps can be related to scattering induced level
shifts. For the 2.5 nm barrier the results are in good agreement with
the measurements reported in Ref.~\onlinecite{HelmPRB1993}. The onset of
absorption around 100 meV is slightly sharper in the experiment, which
may be attributed to less rough interfaces or to the limited
accuracy of the various approximations used for the scattering potentials. 

For SLs the absorption can be understood within the common
miniband picture. For low frequencies intra-miniband processes
dominate, which are easily treated in semiclassical transport
models providing for zero electric field\cite{KtitorovSPS1972,IgnatovPRL1993} 
\begin{equation}
\alpha(\omega)=\frac{\Re\{\sigma(\omega)\}}{c\sqrt{\epsilon}\epsilon_0}
=\frac{\sigma_0}{c\sqrt{\epsilon}\epsilon_0}\frac{1}{(\tau_m\omega)^2+1}\, .
\label{EqalphaMB}
\end{equation} This behavior was experimentally observed in
Refs.~\onlinecite{BrozakPRL1990,TamuraPHB1999}.  Here, $\sigma_0\approx
n_ce^2\tau_m/m_\textrm{SL}$ for large miniband widths. With decreasing
miniband width, the increase of $m_\textrm{SL}$ reduces $\sigma_0$. An even
stronger reduction arises, if the miniband width drops below either $k_BT$ or
the Fermi energy, see Ref.~\onlinecite{WackerPR2002} for  details.\footnote{The
sequential tunneling picture provides similar results for $\sigma(\omega)$
\cite{WackerPR2002}. Thus no major differences are  expected for thick
barriers} For all superlattice structures studied by our NEGF model, we found
good agreement with (\ref{EqalphaMB}) for low frequencies. Some examples are
shown in the inset of Fig.~\ref{FigSLabs}(b). As a further example, the
calculated absorption spectrum at 65 K for the structure  of 
Ref.~\onlinecite{TamuraPHB1999} can be fitted by $\tau_m= 0.16$ps (not shown
here). This is in good agreement with the experimental value of 0.18 ps,
which demonstrates the quality of the NEGF approach.

For higher frequencies, transitions between the minibands can
describe the absorption between 60 and 200 meV very well. See, e.g., the
results of the calculations in Ref.~\onlinecite{HelmPRB1993}, which fully agree with our
more sophisticated NEGF approach. 

We conclude that the absorption of SLs at zero bias can be well described by
the Drude-like miniband conduction result (\ref{EqalphaMB})  for low
frequencies and by common inter-miniband transitions for higher frequencies.
As shown in Fig.~\ref{FigSLabs}(b), the combination of both features evolves
into the bulk FCA (\ref{EqDrude}) if the barrier width becomes small.

\section{From multiple well to superlattice}
\begin{figure}
\includegraphics[width=0.43\columnwidth]{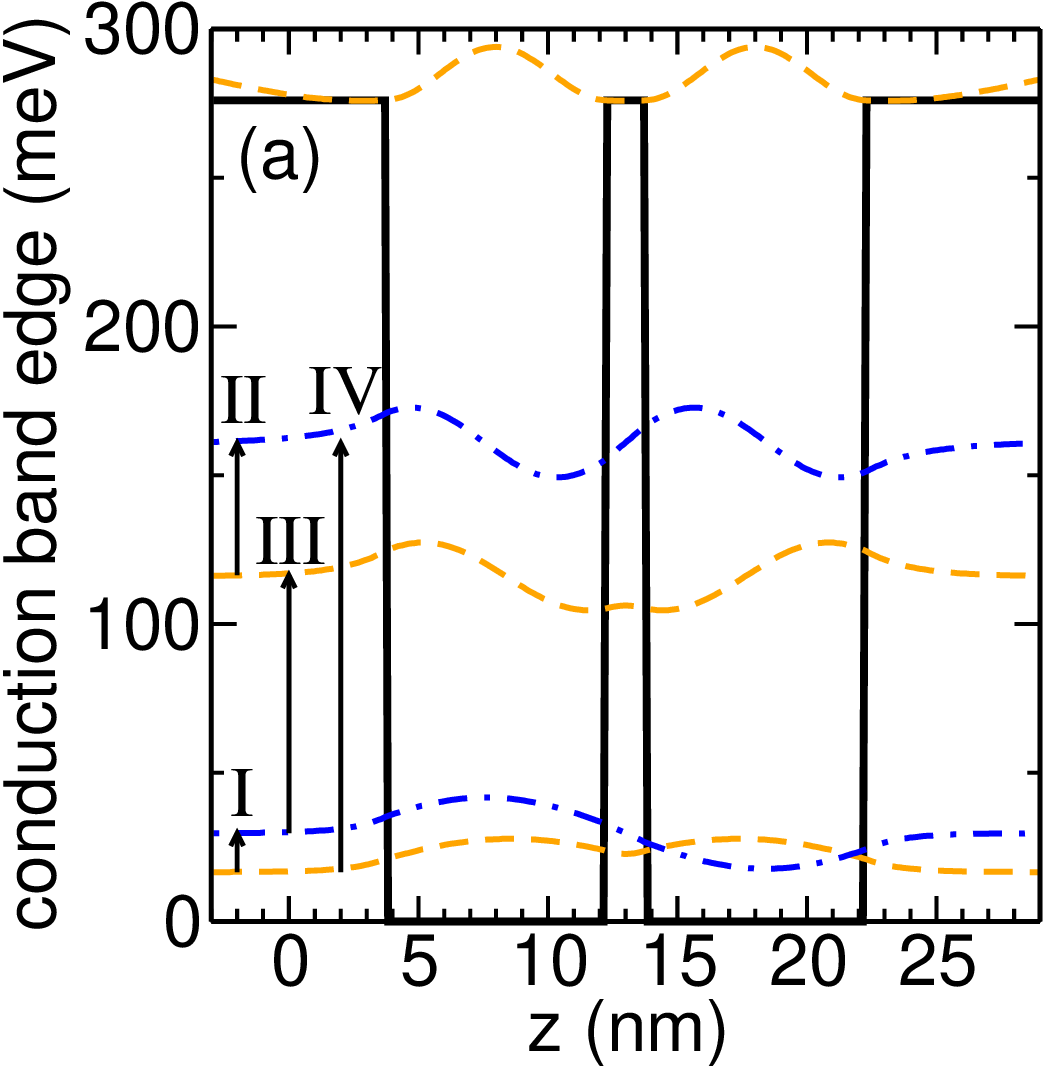}
\hspace{0.04cm}
\includegraphics[width=0.43\columnwidth]{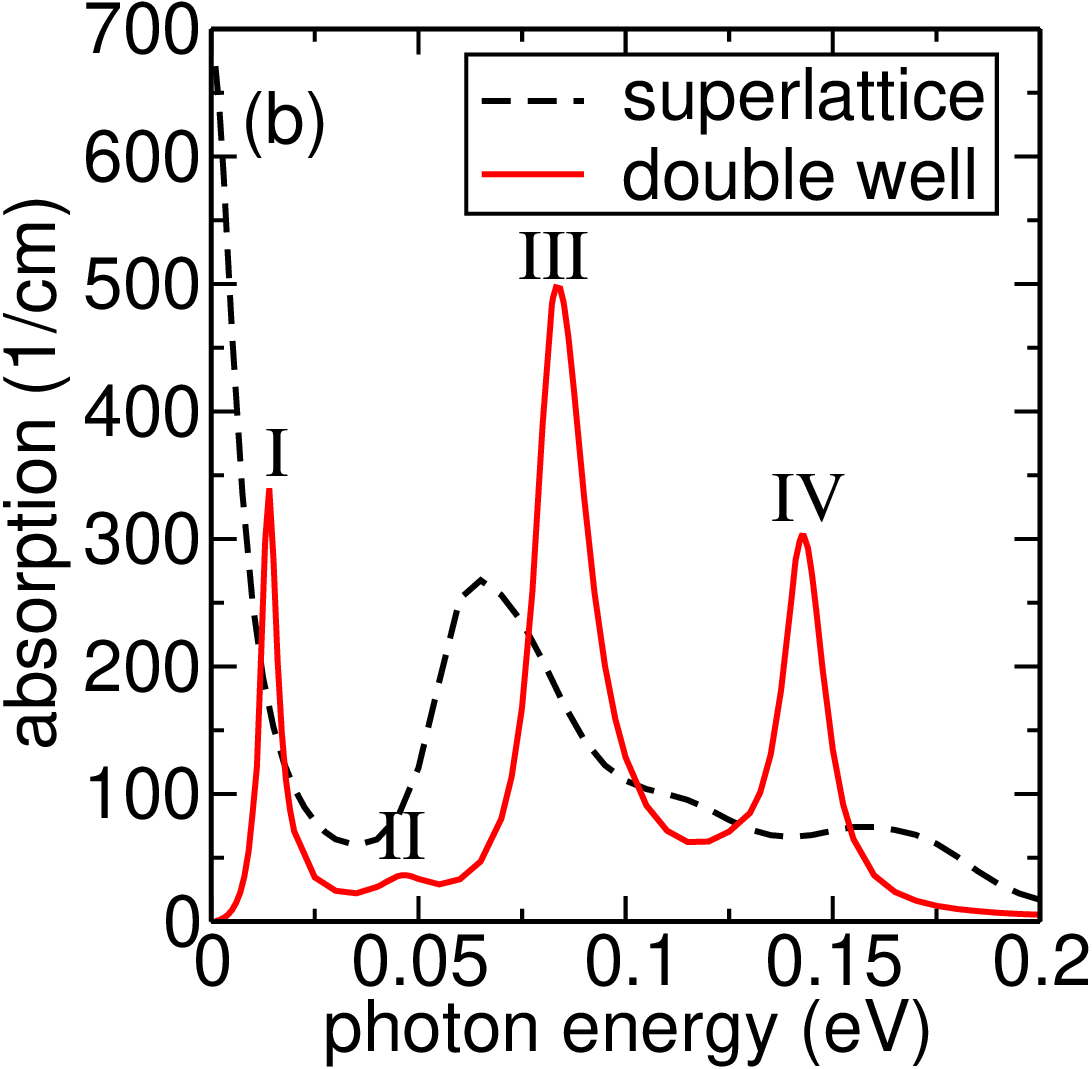}\\
\includegraphics[width=0.9\columnwidth]{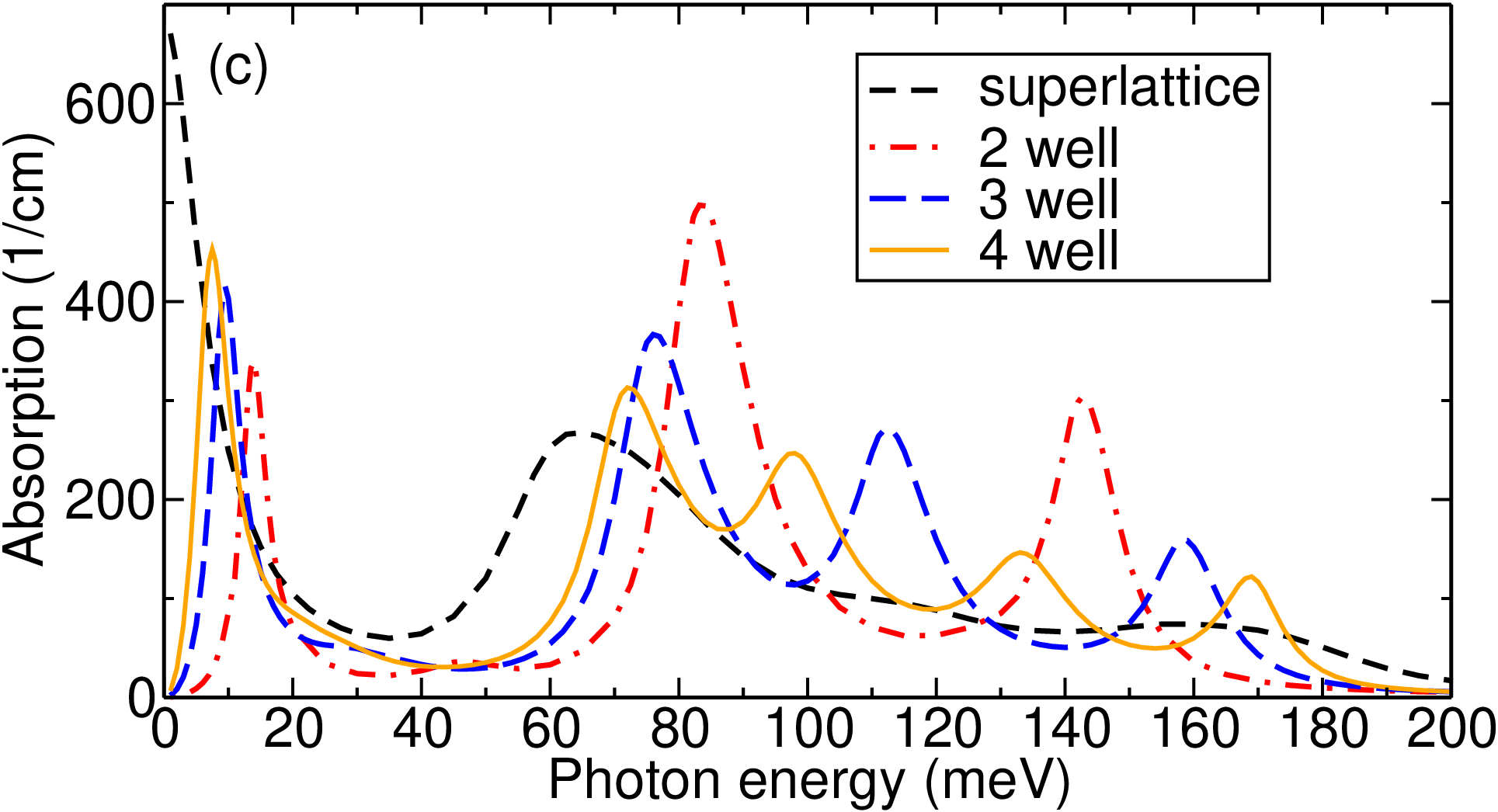}
\caption{(Color online) (a) A double quantum well (well widths 8.5 nm,
barrier width 1.5 nm) with its lowest eigenstates. Dashed and dot-dashed lines
refer to symmetric and antisymmetric states, respectively.  The arrows depict
the transitions associated with the peaks in the absorption spectrum. (b)
Absorption spectrum calculated by the NEGF model for the double quantum well
and the corresponding SL. (c) Evolution of the absorption for 2, 3, and 4
wells with the same parameters as the double well from (a). In order to obtain
absorption in the entire waveguide, it is assumed that the multi-quantum-well structure
is periodically repeated with a separation by a 7.5 nm barrier. All
calculations are done at $T=300$ K.}
\label{FigWellAbs}
\end{figure}

Now we want to study, how the SL absorption arises from the
behavior of systems containing few wells, which show distinct absorption peaks
between discrete levels. Fig.~\ref{FigWellAbs} shows the absorption
for multi-quantum-well structures, as presented in
Fig.~\ref{FigWellAbs}(a) for the case of two wells. Here, all
parameters correspond to the SL with a 1.5 nm barrier
discussed above. For the double well structure, essentially the two
lowest subbands are occupied in thermal equilibrium, and one observes
clear absorption peaks corresponding to the separations between the
subbands, see Fig.~\ref{FigWellAbs}(b). As the dipole matrix element
(\ref{Eqzmatrix}) vanishes for equal parity of the states, not all
possible transitions are visible. The observed peak structure can be
directly described by the standard intersubband expressions
(\ref{EqAlphaIntersubband}). Furthermore, there is
zero absorption in the limit of zero frequency as no dc current along
the structure is possible, compare Eq.~(\ref{EqAlphaSigma}). 

With increasing well numbers, the peaks III and IV of the double well
split up and form the continuous absorption between 60 and 200 meV due
to the transitions between the first and the second SL
miniband, see Fig.~\ref{FigWellAbs}(c). While this is quite expected,
peak I does not show any clear splitting, but shifts to lower
frequencies, approaching the intra-miniband absorption. This behavior
can be understood by a detailed study of the multi-quantum-well
eigenstates. Here, a tight binding model for $N$ wells with next neighbor
coupling $T_1$ shows (see the appendix \ref{SecAppendix} for details): (i) There are
$N$ eigenstates, labeled by an index $\nu$ according to their energy
$E_\nu$.  Here $E_{\nu+1}-E_\nu$ is of the order of $4|T_1|/N$. (ii)
The matrix element $z_{\mu\nu}$ from Eq.~(\ref{Eqzmatrix}) is small
unless for neighboring states, i.e. $\mu=\nu\pm 1$. Thus, the
transitions between neighboring states dominate, explaining  the
strong absorption around $\hbar\omega\approx 4|T_1|/N$ visible in
Fig.~\ref{FigWellAbs}(c), where $ 4|T_1|$ essentially corresponds to
the miniband width of the infinite structure. Together with a tail at
higher frequencies due to broadening of these transitions this
explains the appearance of the Drude-like miniband absorption for the
SL in the limit of large $N$. For $\omega=0$ the evolution
is not smooth as any finite sequence of quantum wells has a zero dc
conductivity in contrast to an infinite SL and thus the
absorption must vanish according to Eq.~(\ref{EqAlphaSigma}).

\section{The integrated absorption}

Summing over all possible intersubband transitions 
(\ref{EqAlphaIntersubband}), we obtain the total absorption 
$\alpha_\mathrm{IS}(\omega)=\sum_{\mu\nu}\alpha_{\mu\to\nu}(\omega)
\Theta(E_{\nu}-E_{\mu})$. Here the discrete index $\nu$ runs over 
all (infinitely many) eigenstates of the heterostructure of finite length, 
including states which correspond to unbounded states with energies far 
above the barrier potential. Integrating over all frequencies provides
\begin{equation}\begin{split}
&\int_0^\infty\d \omega\; \alpha_\mathrm{IS}(\omega)\\
&=\sum_{\nu, \mu}\frac{\pi e^2|z_{\mu,\nu}|^2(E_{\nu}-E_{\mu})
(n_{\mu}-n_{\nu})} {L_zc\epsilon_0\sqrt{\epsilon}\hbar^2}
\Theta(E_{\nu}-E_{\mu})\\
&=\sum_{\mu, \nu}\frac{\pi e^2|z_{\mu,\nu}|^2(E_{\nu}-E_\mu)
n_\mu} {L_zc\epsilon_0\sqrt{\epsilon}\hbar^2}
\label{EqalphaIS}
\end{split}\end{equation}
under the assumption $E_{\nu}-E_{\mu}\gg \Gamma$ -- otherwise the counter-rotating
terms become of relevance, which had been neglected here. 
In appendix \ref{SecAppGFT}
we show that the same integral relation is more generally obtained for
arbitrary level spacings $E_{\nu}-E_{\mu}$
within our NEGF model, which also covers dispersive
gain.\cite{TerazziNaturePhys2007,WackerNaturePhys2007}

Following Ref.~\onlinecite{PeetersPRB1993}, Eq.~(\ref{EqalphaIS}) can
be simplified by the Thomas-Reiche-Kuhn sum
rule\cite{KuhnZP1925,*ReicheZP1925} (also called f-sum
rule\cite{MahanBook1990}) which reads for a parabolic band with effective
mass $m_c$ 
\[
\sum_\nu \frac{2m_c(E_\nu-E_\mu)}{\hbar^2}|z_{\mu\nu}|^2=1
\]
and provides the integrated absorption
\begin{equation}
\int_0^\infty\d \omega \; \alpha_\mathrm{IS}(\omega)
=n_\mathrm{av}\frac{\pi e^2} {2m_cc\sqrt{\epsilon}\epsilon_0}
\label{EqIntIntersubband}
\end{equation}
where $n_\mathrm{av}=\sum_\mu  n_{\mu}/L_z$ is the average three-dimensional
carrier density in the waveguide.

For a bulk semiconductor, the free carrier absorption
(\ref{EqDrude}) provides after integration over energy
\begin{equation}
\int_0^\infty\d \omega \;  \alpha_\mathrm{FCA}(\omega)
= n_c\frac{\pi e^2}{2m_cc\sqrt{\epsilon}\epsilon_0}
\label{EqIntFCA}
\end{equation}  which fully agrees with the intersubband result
(\ref{EqIntIntersubband}) for equal total densities $n_c=n_\mathrm{av}$.  Thus
the total FCA in a bulk semiconductor equals the total intersubband
absorption within the conduction band for a finite heterostructure of  finite
length, which shows the direct relation between these. More  generally,
Eqs.~(\ref{EqIntIntersubband},\ref{EqIntFCA}) establish a general rule for the
integrated absorption within the conduction band of a semiconductor under
conditions, where the approximation of a constant effective mass is justified.
In this context superlattices appear as an intermediate case, where  the
inter-miniband absorption and the  Drude-like intra-miniband absorption add up
to the full result.\cite{PeetersPRB1993}

Our numerical data in Fig.~\ref{FigSLabs}(b) exhibit the integrated absorption
$\hbar\int\d \omega\; \alpha(\omega)=25.7(\pm 0.3)\mathrm{eV}/\mathrm{cm}$ for all
curves.
The data from Fig.~\ref{FigWellAbs} provide 
$\hbar\int\d \omega\; \alpha(\omega)=\frac{N}{N+0.6}
25.6(\pm 0.2)\textrm{ eV}/\mathrm{cm}$, where the additional factor takes into
account the undoped region of 6 nm between adjacent multiple quantum wells ($N$
is the number of well/barrier combinations with a length of 10 nm each). 
These values are slightly below the value of $27.3\textrm{ eV}/\mathrm{cm}$ 
given by Eq.~(\ref{EqIntFCA}) using the GaAs effective mass. This minor
discrepancy of less than 7\% 
can be easily attributed to some absorption at higher 
frequencies and the impact
of the barrier material with a larger mass.\footnote{Indeed we found a
  consistent change with the barrier thickness: The maximal value
  of $26\textrm{ eV}/\mathrm{cm}$ was obtained for the  0.5 nm barrier and the 
  minimal value of $25.4\textrm{ eV}/\mathrm{cm}$ for the 3.5 nm barrier in
  Fig.~\ref{FigSLabs}(b).}
We conclude, that the absorption obtained by our NEGF code is in 
excellent agreement with the rule (\ref{EqIntIntersubband}).

More generally the effect of the semiconductor heterostructure can be understood
as shifting the absorption strength within the frequency space, as explicitly
demonstrated by our calculations. This perception has
actually been used in the design of QCL
structures, where the unavoidable free carrier absorption is 
deflected from the frequency
region of operation by a proper choice of heterostructures
\cite{FaistPrivCom}, see, e.g., Ref.~\onlinecite{WaltherAPL2006}.

\section{Conclusion} We demonstrated how the common bulk free carrier
absorption evolves from standard intersubband absorption in heterostructures
for electromagnetic waves with an electric field pointing in growth direction.
Here the well-studied SL absorption constitutes an intermediate case, which
can be entirely understood on the basis of common intersubband absorption
processes in the limit of a growing number of quantum wells. For decreasing SL
barrier width the combination of inter- and intra-miniband absorption evolves
into the standard FCA of the bulk crystal. This behavior reflects a
redistribution of absorption strength, while the integrated absorption is
constant.
The most relevant consequence is
that there is no need to bother about any additional FCA-related
absorption processes, provided all intersubband transitions are properly taken
into account. A consistency check for the calculated  gain/absorption spectrum
is whether Eq.~(\ref{EqAlphaSigma}) is satisfied in the low frequency limit
and the integrated absorption matches Eqs.~(\ref{EqIntIntersubband},\ref{EqIntFCA}).


\begin{acknowledgments}
We thank J. Faist for helpful discussions.
Financial support from the Swedish Research Council (VR) and the French ANR 
agency (ROOTS project) is gratefully acknowledged.
\end{acknowledgments}

\appendix

\section{Analytical calculation for coupled wells}\label{SecAppendix}
We consider a multi-quantum-well structure with $N$ wells centered at
$z=nd$, where $n=1,2,\ldots N$. The ground state of the isolated  well
$n$ has the wavefunction $\Psi_g(z-nd)$ and the energy
$E_g$. Restricting to a nearest neighbor coupling $T_1$ (which is
negative for the lowest subband), the eigenenergies are
\begin{equation}
E_{\nu}=E_g+2T_1\cos\left(\frac{\nu \pi}{N+1}\right)\quad \mbox{for}\,
\nu=1,2,\ldots N \label{EqEnergyMQW}
\end{equation}
and the eigenstates read
$\varphi_{\nu}(z)=\sum_na_n^{(\nu)}\Psi_g(z-nd)$ with
\begin{equation*}
a_n^{(\nu)}=\sqrt{\frac{2}{N+1}}\sin\left(\frac{\nu \pi n}{N+1}\right)\, .
\end{equation*}
If the overlap between the states in different wells is
negligible, i.e. $\int\d z \Psi_g(z-n'd)z\Psi_g(z-nd)\approx
nd\delta_{nn'}$, we find $z_{\mu\nu}=\sum_n
nd a_n^{(\mu)}a_n^{(\nu)}$, which can be directly evaluated. If
$\nu-\mu$ is even we find $z_{\mu\nu}=\delta_{\mu,\nu}(N+1)d/2$ as 
both states have the same parity with respect to
$z=(N+1)d/2$. For odd $\nu-\mu$, some algebra yields
\begin{equation*}
z_{\mu\nu}=\frac{d}{2(N+1)}\left[
\frac{1}{\sin^2\left(\frac{(\mu+\nu)\pi}{2(N+1)}\right)}
-\frac{1}{\sin^2\left(\frac{(\mu-\nu)\pi}{2(N+1)}\right)}
\right]
\end{equation*}
For $\mu\neq\nu$ we thus have:
\begin{alignat*}{2}
z_{\mu\nu}&=0 &\quad & \mbox{for even}\, (\nu-\mu)\\  
z_{\mu\nu}&\sim -\frac{2(N+1)d}{\pi^2(\mu-\nu)^2}
& \quad&\mbox{for odd and small}\, (\nu-\mu)\\
z_{\mu\nu}&=O\left\{\frac{d}{N+1}\right\} &\quad &\mbox{for odd and large}\, (\nu-\mu)
\end{alignat*}
As the square of $z_{\mu\nu}$ enters the absorption
(\ref{EqAlphaIntersubband}), it becomes clear that the transitions
with $\nu=\mu\pm 1$ highly dominate the absorption spectrum. The energy
difference of the corresponding states (\ref{EqEnergyMQW}) 
for these transitions is less
than  $2|T_1|\pi/(N+1)$ with an average of approximately $4|T_1|/N$.

\section{Total absorption with the Green's function model}\label{SecAppGFT}
Here we refer to the formulation of our NEGF model
as outlined in Ref.~\onlinecite{WackerSPIE2009}. Here gain is
  evaluated within linear response around the stationary state characterized by
  the Green's functions $\tilde{G}_{\mu\nu}({\bf k},E)$. In order to simplify
  the analysis, nondiagonal  $\tilde{G}_{\mu\nu}({\bf k},E)$ are neglected
  here -- they are, however, fully included in our numerical 
implementation. Then the absorption resulting from the pair of states $\mu,\nu$ can be written as
\begin{widetext}
\begin{equation}\begin{split}
\alpha_{\mu\nu}(\omega)=\frac{e^2(E_\nu-E_\mu)|z_{\mu\nu}|^2}
{cL_z\hbar\epsilon_0\sqrt{\epsilon}}
\frac{2}{A}\sum_{\bf k}
\int\frac{\d E}{2\pi}
\Re\Big\{&
\tilde{G}_{\nu\nu}^{\text{ret}}({\bf k},E+\hbar \omega)
\tilde{G}^{<}_{\mu\mu}({\bf k},E)
+\tilde{G}^{<}_{\nu\nu}({\bf k},E+\hbar \omega)
\tilde{G}^{\text{adv}}_{\mu\mu}({\bf k},E)\\
&-\tilde{G}_{\mu\mu}^{\text{ret}}({\bf k},E+\hbar \omega)
\tilde{G}^{<}_{\nu\nu}({\bf k},E)
-\tilde{G}^{<}_{\mu\mu}({\bf k},E+\hbar \omega)
\tilde{G}^{\text{adv}}_{\nu\nu}({\bf k},E)\Big\}
\end{split}\end{equation}
which is essentially the last equation of the appendix in 
Ref.~\onlinecite{WackerSPIE2009}
with the counter-rotating term added. Inserting the spectral function\cite{HaugJauhoBook1996} 
$A_\nu({\bf k},E)=\mp 2\Im\{\tilde{G}^{\text{ret/adv}}_{\nu,\nu}({\bf k},E)\}$
and its occupied part\footnote{In thermal equilibrium we have
$A^\textrm{occ}_\nu({\bf k},E)=n_F(E)A_\nu({\bf k},E)$, where $n_F(E)$ is the
Fermi-Dirac distribution.}
$A^\textrm{occ}_\nu({\bf k},E)=-\imai\tilde{G}^{<}_{\nu\nu}({\bf k},E)$,
which is assumed to be real, we find 
\begin{equation}\begin{split}
\alpha_{\mu\nu}(\omega)=\frac{e^2(E_\nu-E_\mu)|z_{\mu\nu}|^2}
{2cL_z\hbar\epsilon_0\sqrt{\epsilon}}
\frac{2}{A}\sum_{\bf k}
\int\frac{\d E}{2\pi}
\Big[&
 A^\textrm{occ}_\mu({\bf k},E)A_\nu({\bf k},E+\hbar \omega)
- A^\textrm{occ}_\nu({\bf k},E)A_\mu({\bf k},E-\hbar \omega)\\
&+A^\textrm{occ}_\mu({\bf k},E)A_\nu({\bf k},E-\hbar \omega)
-A^\textrm{occ}_\nu({\bf k},E)A_\mu({\bf k},E+\hbar \omega)
\Big]
\label{EqalphaGFT}
\end{split}\end{equation}
The terms
$A^\textrm{occ}_\mu({\bf k},E)A_\nu({\bf k},E+\hbar \omega)
-A^\textrm{occ}_\nu({\bf k},E)A_\mu({\bf k},E-\hbar \omega)$ provide
the physical origin of dispersive gain as sketched in
Refs.~\onlinecite{WackerNaturePhys2007,WackerSPIE2009}. The signs of the
counter-rotating terms 
$A^\textrm{occ}_\mu({\bf k},E)A_\nu({\bf k},E-\hbar \omega)
-A^\textrm{occ}_\nu({\bf k},E)A_\mu({\bf k},E+\hbar \omega)$
seem to contradict our intuition, as the first one appears to relate to
emission and the second to absorption. However, in this formulation
the sign is defined via the difference in energy between the initial and the 
final state, where only one a specific combination is used in the 
prefactor $(E_\nu-E_\mu)$.

Using the general relations
\[
\int_0^\infty  \d \omega
\left[A_\mu({\bf k},E+\hbar \omega)+A_\mu({\bf k},E-\hbar \omega)\right]
=\frac{1}{\hbar}\int_{-\infty}^\infty \d E' A_\mu({\bf k},E')=2\pi/\hbar
\quad\mbox{and}\quad
\frac{2}{A}\sum_{\bf k}\int\frac{\d E}{2\pi} A^\textrm{occ}_\mu({\bf k},E)=n_\mu
\] 
integration of the terms from Eq.~(\ref{EqalphaGFT}) over frequency provides
\begin{equation}
\int_0^{\infty}\d\omega \alpha_{\mu\nu}(\omega)=
\frac{\pi e^2|z_{\mu,\nu}|^2(E_{\nu}-E_{\mu})(n_{
\mu}-n_{\nu})} {L_zc\epsilon_0\sqrt{\epsilon}\hbar^2}
\end{equation}
so that the sum over all different pairs $(\mu,\nu)$ equals 
the second line of Eq.~(\ref{EqalphaIS}).
Thus the integrated absorption (\ref{EqIntIntersubband}) also holds for 
the more involved absorption terms (\ref{EqalphaGFT}) of the NEGF model 
which include the dispersive gain. 
\end{widetext}


%

\end{document}